\documentclass[12pt]{article}
\usepackage{fullpage}
\usepackage{amsmath}
\usepackage{bm}
\usepackage{amssymb}
\usepackage{booktabs}
\usepackage{setspace}
\usepackage{color}

\everydisplay{\abovedisplayshortskip\belowdisplayshortskip}%
\makeatletter \oddsidemargin  -.1in \evensidemargin -.1in
\title{\bf Acceptance sampling plans for inverse Weibull distribution based on truncated life test}
\author{ Sukhdev Singh and Yogesh Mani Tripathi \footnote{Corresponding author. E-mail address: yogesh@iitp.ac.in}\\
\small Department of Mathematics, Indian Institute of Technology Patna,  India}
\date{}
\begin{document}
\begin{singlespace}
\maketitle
\end{singlespace}
\bibliographystyle{agsm}
\setcounter{page}{1}
\setcounter{equation}{0}
\setcounter{figure}{0}
\setcounter{table}{0}
\begin{abstract}
\noindent In this paper, we develop double acceptance sampling plan and group acceptance sampling plan for an inverse Weibull distribution based on a truncated life test. We consider the median lifetime of the test units as a quality parameter and obtain the design parameters such as sample size and acceptance number. These plans are obtained under the consumer's risk and the producer's risk simultaneously involved at a certain confidence level. We present a simulation study to support the proposed methods and a comparison between single and double acceptance sampling plans is made. A real data set is also analyzed to illustrate the implementation of the proposed sampling plans. Further, the situation under which the proposed samplings plans can also be used for other percentiles points is discussed. Finally a conclusion is presented.\\

\end{abstract}
\section{\bf {Introduction}}
\label{sec:1}
In many statistical analysis, lifetime of a product is treated as an important quality characteristic. Acceptance sampling plans render one such assurance on product quality. Such plans have attracted much attention in the area of quality control and life testing studies. In general, sampling plans associated with some life test are conducted under various restrictions on experimental time and experimental cost. An important aspect of acceptance sampling is to make the decision either to accept or to reject the concerned lot of products based on quality characteristic observed from the given life test. This decision making process commonly involves both producer's risk as well as consumer's risk. Notice that probability $\beta$ of accepting a bad lot is termed as consumer's risk whereas probability $\alpha$ of rejecting a good lot is termed as producer's risk. In an acceptance sampling plan, when quality characteristic of experimental units is measured on a numerical scale such as failure times then the corresponding plans are called variable sampling plan. On the other hand, if quality characteristic is measured on the basis of number of failures in the observed sample taken from the lot then it is known as attribute sampling plan. In this work, we are interested in developing attribute sampling plan based on a truncated life test in which the experiment under consideration is terminated at a pre-assigned time $t$. Consider a situation in which a random sample of $n$ units taken from a lot is placed on a life test for $t$ units of time. During the $t$ units of time if more than a specified number $c$ (acceptance number) of units fail then the consumer accept that lot, otherwise, the lot is rejected. Such type of sampling plan is called single acceptance sampling plan, where $n$ and $c$ are called the design parameters of the plan. In literature, single acceptance sampling plans based on truncated life test have been proposed for many lifetime distributions. Epstein \cite{Epstein-1954} introduced such plans for exponential distribution. Goode and Kao \cite{Goode-Kao-1961} reported acceptance sampling plans for Weibull distribution. Plans for normal and lognormal distributions were obtained by Gupta \cite{Gupta-1962}. For some more recent works on single acceptance sampling, one may refer to Rosaiah and Kantam \cite{Rosaiah-Kantam-2005}, Rosaiah {\sl{et al.}} \cite{Rosaiah-Kantam-Kumar-2006}, Tsai and Wu \cite{Tsai-Wu-2006}, Balakrishnan {\sl{et al.}} \cite{Balakrishnan-Leiva-Lopez-2007} and Aslam {\sl{et al.}} \cite{Aslam-Kundu-Ahmed-2010}.

Recall that, in single acceptance sampling plan based on the truncated life test, a decision is made as either to accept the proposed lot or to reject the lot. A more generalized sampling plan known as double acceptance sampling plan has also found wide applications in several statistical analysis including quality control and reliability studies. Aslam and Jun \cite{Aslam-Jun-2010} reported double acceptance sampling plans for a generalized log-logistic distribution. In their work, authors have developed the corresponding plans under zero one failure scheme, in which a lot is accepted if no failures are observed from the first sample and it is rejected if two or more failures are observed. Further, in case of one failure, a second sample is drawn and is tested exactly for the same time period as the first sample. One may also refer to Aslam {\sl{et al.}} \cite{Aslam-Jun-Ahmed-2011} for some more work on double acceptance sampling plans for Birnbaum-Saunders distribution. In many practical situations, units are tested in groups where each group contain equal number of units which are tested simultaneously under identical conditions. The corresponding plan is referred to as a group acceptance sampling plan. In life testing such plans are very much useful in the situations when the units on test are highly reliable but of comparatively less cost. Many researchers have obtained group acceptance sampling plan for various lifetime distributions and one may refer to  Aslam {\sl{et al.}} \cite{Aslam-Jun-Ahmed-2009} for gamma distribution, Aslam and Jun \cite{Aslam-Junn-2009} for Weibull distribution, Aslam and Jun \cite{Aslam-Jun-2009} for log-logistic distribution, Aslam {\sl{et al.}} \cite{Aslam-Jun-Ahmed-2011} for Birnbaum-Saunders distribution, and Singh {\sl{et al.}} \cite{singh2015sampling} for generalized inverted exponential distribution.

In this paper, we develop double acceptance sampling plan and group acceptance sampling plan based on the truncated life test. The rest of this paper is organized as follows. In Section \ref{sec:2}, some basic properties of the inverse Weibull distribution are discussed. Double acceptance sampling plan are proposed in Section \ref{sec:3}. A numerical comparison between single and double acceptance sampling plans and an illustrative example are also presented in this section. In Section \ref{sec:4}, group acceptance sampling plans are reported along with an example. Section \ref{sec:5}, deals with the selection criterion for the shape parameter in practical implementations. In Section \ref{sec:6}, sampling plans for 100$p$th percentile points are discussed. Finally, a conclusion is given in Section \ref{sec:7}.

\section{\bf {Inverse Weibull distribution}}
\label{sec:2}
A two-parameter inverse Weibull distribution was first introduced in literature by Killer and Kamath \cite{Keller-Kanath-1982} as a suitable model to describe degradation phenomena of mechanical components of diesel engines. It has found widespread applications in various areas of reliability analysis. The probability density function (PDF) of this distribution is given by
\begin{eqnarray}
\label{2.1}
f(t; \gamma, \lambda) &=& \gamma \lambda  t^{-(\gamma+1)}e^{-\frac{\lambda}{t^{\gamma}}}, ~~ 0<t<\infty, \,\, \lambda,\gamma > 0,
\end{eqnarray}
where $\gamma$ is a shape parameter which governs the shape of the distribution and $\lambda$ is a scale parameter which governs the dispersion of the distribution. Furthermore depending upon the values of $\gamma$, the corresponding hazard function can be monotonic decreasing or unimodal.  We refer to Kundu and Howlader \cite{Kundu-Howlader-2010} for some more discussions on various properties of this distribution. The works of Erto and Rapone \cite{Erto-Rapone-1984}, Calabria and Pulcini \cite{Calabria-Pulcini-1994} and Murthy {\sl{et al.}} \cite{Murthy-2004} suggest that inverse Weibull distribution can be treated as a lifetime model to various failure lifetime data. For examples, the physical process reported by Erto and Rapone leads to an inverse Weibull distribution. Authors also showed that this distribution provides reasonably good fit to the survival data such as the times to breakdown of an insulating fluid subject to the action of a constant tension. One may also refer to Nelson \cite{Nelson-1982}, Erto \cite{Erto-1989}, Kim {\sl{et al.}} \cite{Kim-Lee-Kang-2012} and Sultan {\sl{et al.}} \cite{Sultan-Alsadat-Kundu-2013} for many interesting statistical inferences on inverse Weibull distribution. We denote this distribution as $IW(\gamma, \lambda)$. In the present work, we construct acceptance sampling plan for the 100$p$th percentile $\theta_{p} = F^{-1}(p; \lambda, \gamma)$ points. Observe that $\theta_{p}$ is given by
\begin{eqnarray}
\label{2.2}
\theta_{p}&=& \Big(\frac{-\lambda}{log p}\Big)^{\frac{1}{\gamma}}.
\end{eqnarray}
As a consequence the median of the distribution is given by
\begin{eqnarray}
\label{2.3}
m &=& \Big(\frac{\lambda}{log 2}\Big)^{\frac{1}{\gamma}}.
\end{eqnarray}
For further consideration we treat the median lifetime $m$ of the test unit as a quality parameter and accordingly, construct acceptance sampling plans. We mention that a discussion of proposed acceptance sampling plans for other percentile points is presented in Section \ref{sec:6}.
\section{\bf{Design of double acceptance sampling plan}}
\label{sec:3}
Consider an experimental situation where lifetimes of the test units follow an $IW(\gamma, \lambda)$ distribution and a producer claims that the specified median lifetime of the units is $m_0$. We are interested in making inference whether the actual median lifetime, $m$ of an unit is larger than a prescribed lifetime $m_{0}$. The usual practice is to take a random sample from that lot and then perform a truncated life test for, say, $t_{0}$ units of time, where $t_{0}$ can be some multiple of $m_{0}$. That is for any positive constant $a$, we may take $t_{0} = am_{0}$.  Then a lot under investigation will be accepted if there is enough evidence that $m \geq m_{0}$, at certain level of producer's risk $\alpha$ and consumer's risk $\beta$. Accordingly, we suggest the following double acceptance sampling plan.\\

\noindent
(1) Perform a life test for $t_{0}$ units of time on a random sample of size $n_{1}$, drawn from a lot\\$~~~~~$ under investigation. \\
(2) Accept the lot if $c_{1}$ or less number of units fail during the test.  If $c_{2}+1$ or more number\\$~~~~~$ of units fail then terminate the experiment and reject the lot. Next, consider the case where\\$~~~~~$  the number of failed units lie between $c_{1}+1$ and $c_{2}$. Draw a second random sample of size\\$~~~~~$ $n_{2}$ and subject them to test for the same duration $t_{0}$.\\
(3) Accept the lot if at most $c_{2}$ number of units fail from the two samples, otherwise, reject\\$~~~~~$ the lot.\\

The proposed double acceptance sampling is formulated using design parameters $(n_{1}, n_{2}, c_{1}, c_{2})$, where $c_{1} < c_{2}$ and $n_{2} \leq n_{1}$. Note that the single acceptance sampling plan is a special case of double acceptance sampling plan when $c_{1} = c_{2} =c$, whereas, zero one failure scheme is a particular case of the said plan for the choice $c_{1} = 0$ and $c_{2} = 1$. Now observe that the probability of accepting the lot under the proposed sampling plan is given by
\begin{eqnarray}
\label{3.1}
P_{a}(p) &=& \sum_{i=0}^{c_{1}}{n_{1} \choose i}p^{i}(1-p)^{n_{1}-i} + \sum_{j=c_{1}+1}^{c_{2}}{n_{1}\choose j}p^{j}(1-p)^{n_{1}-j}\Bigg[\sum_{i=0}^{c_{2}-j}{n_{2}\choose i}p^{i}(1-p)^{n_{2}-i}\Bigg],
\end{eqnarray}
where $p=F(t_{0}; \gamma, \lambda)$ is the probability that a test unit fails before the termination time point $t_{0}$ such that
\begin{eqnarray*}
p &=& e^{-\frac{\lambda}{{t^{\gamma}_{0}}}}. 
\end{eqnarray*}
However using $t_{0}=am_{0}$ and the equation (\ref{2.3}), $p$ can be written as
\begin{eqnarray}
\label{3.2}
p &=& e^{-{(\frac{m}{m_{0}})}^{\gamma}\,log 2\, a^{-\gamma}}.
\end{eqnarray}
Thus for a given value of $a, m, m_{0}$ and the shape parameter $\gamma$, the corresponding $p$ can be computed. Also observe that $p$ is independent of the scale parameter $\lambda$. However, median lifetime of the product accounts the information of both the shape and scale parameter, see equation (\ref{2.3}).
Further we consider the ratio $\frac{m}{m_{0}}$ as a quality level. In general, from a consumer perspective probability of accepting a lot should be smaller than the probability of accepting a bad lot when $m=m_{0}$. On the other hand, producer requires that the probability of rejecting a lot should be smaller than the probability of rejecting a good lot when $m>m_{0}$. So, for a given quality level $\frac{m}{m_{0}}$, at consumer's risk $\beta$ and producer's risk $\alpha$  together with $a$, design parameters $(n_{1}, n_{2}, c_{1}, c_{2})$ can be obtained by solving the following two inequalities simultaneously
\begin{eqnarray}
\label{3.3}
P_{a}\Big(p_{1} \mid \frac{m}{m_{0}} = r_{1}\Big) & \leq& \beta ,\\
\label{3.4}
P_{a}\Big(p_{2} \mid \frac{m}{m_{0}} = r_{2}\Big) &\geq& 1-\alpha,
\end{eqnarray}
where $p_{1}$ is the probability that a test unit fails before the termination time $t_{0}$ when $r_{1}$ is the quality level corresponding to consumer's risk $\beta$ and $p_{2}$ is the probability that a test unit fails before the termination time $t_{0}$ when $r_{2}$ is the quality level corresponding to producer's risk $\alpha$. Observe that, multiple solutions $(n_{1}, n_{2}, c_{1}, c_{2})$ may satisfy inequalities (\ref{3.3}) and (\ref{3.4}). Here we are interested to find the minimum sample size that satisfy the above mentioned two inequalities. Notice that the minimum average sample number $(ASN)$ required to make a decision to accept or to reject the lot for the proposed double acceptance sampling plan is given by
\begin{eqnarray*}
ASN(p) &=& n_{1}P_{d1}(p)+(n_{1}+n_{2})(1-P_{d1}(p)),
\end{eqnarray*}
where $P_{d1}(p)$ is the probability that a decision is taken from the first sample and it is obtained as
\begin{eqnarray*}
P_{d1}(p) &=& 1-\sum_{i=c_{1}+1}^{c_{2}}{n_{1} \choose i}p^{i}(1-p)^{n_{1}-i}.
\end{eqnarray*}
Consequently, the desired design parameters can be obtained by solving the optimization problem as defined below:\\
Minimize $~~ASN(p_{1}) = n_{1}P_{d1}(p_1)+(n_{1}+n_{2})(1-P_{d1}(p_1))$\\
Subject to\\
$~~~~~~~~~~~~~~~~P_{a}(p_{1} \mid \frac{m}{m_{0}} = r_{1}) \leq \beta $,\\
$~~~~~~~~~~~~~~~~P_{a}(p_{2} \mid \frac{m}{m_{0}} = r_{2}) \geq 1-\alpha$,\\
$~~~~~~~~~~~~~~~~ 1\leq n_{2}\leq n_{1}$,\\
$~~~~~~~~~~~~~~~~n_{1}, n_{2} :$ integers.\\

The design parameters satisfying the above optimization problem are reported in Table \ref{table:1} for $\gamma=0.75$. We mention that three different levels of $a$ such as $0.5, 0.7, 1.0$, four different levels of consumer's risk $\beta$, namely 0.25, 0.10, 0.05, 0.01 and producer's risk $\alpha$ as $0.05$ are taken into consideration. Further the quality level at the consumer's risk $(r_{1})$ is taken as 1, while quality level at the producer's risk $(r_{2})$ are considered as $(\frac{m}{m_{0}} = 2, 3, 4, 5, 6)$. The average sample number $(ASN)$ and the lot acceptance probability $(p_{\alpha})$ at producer's risk with quality level $r_{2}$ are also reported in the table. Tabulated values indicate that with the decrease in consumer's risk $\beta$, the sample size and acceptance number tend to increase. Also, when quality level is allowed to increase with fixed $\beta$, the sample size and the acceptance number tend to decrease. However, this behaviour does not hold true for all quality levels. For example, when $\beta=0.25$ and $a=0.5$, then at quality level $r_{2}=4$, sample size is $(5,4)$ with acceptance constant $(0,1)$. In this case, as the quality level increases sample size and acceptance number remain the same. Influence of time termination multiplier $a$ on the acceptance sampling is also observed. As the value of $a$ increases at a lower quality level $r_{2} \leq 3$, sample size and acceptance number also increases but at higher quality level $r_{2} \geq 4$, sample size and acceptance number start decreasing. Next, Table \ref{table:2}, report design parameters, $ASN$ and lot acceptance probability at producer's risk for $\gamma=1.25$. In fact, we observed that increase in the value of shape parameter leads to smaller sample size. In this case, increase in time termination multiplier $a$ results in a sharp decrease in sample size. Furthermore, a comparison between single and double acceptance sampling plans is presented in Table \ref{table:3}. Recall that smaller sample sizes are more economical from practical applications viewpoint. It is consequently observed from the Table \ref{table:3}, that the sample size obtained using double acceptance sampling plans are smaller than the sample size obtained using single acceptance sampling plans. This holds true for almost all the cases reported in the table. However, the sample size obtained using single acceptance sampling plan is smaller in case when no failure is allowed, that is, when $c=0$. Such a plan is called zero acceptance sampling plan.\\

\noindent \textbf{Example 1:}
Suppose that a producer submits a lot of units and claims that specified lifetime of the units is 2000 hours, here assume that the lifetime of the units follow an inverse Weibull distribution with shape parameter 0.75. Further consider the consumer's risk 10\%, when the true median life of the units is 2000 hours and the producer's risk 5\%, when the true median life of the units is 4000 hours. Now we are interested in the design parameters to implement double acceptance sampling plan when an experimenter would like to run life test experiment for 1000 hours. Notice that, in this case we have $\gamma=0.75$, $m_{0}=2000$ hours, $a=0.5$,  $\beta=0.10$, $r_{1}=1$, $\alpha=0.05$  and $r_{2}=2$. Subsequently, the design parameters from Table \ref{table:1} can be obtained as  $(c_{1}, c_{2}) = (7, 11)$ and $(n_{1}, n_{2}) = (39, 12)$ with $ASN$ 43.43. The consequence of this observation is that a consumer will take a random sample of size $n_{1}=39$ units from proposed lot and then will subject them to a life test for 1000 hours. During the experiment, if at most 7 units fail then that lot will be accepted and if more than 11 units fail then that lot will be rejected. If number of failed units lie between 8 and 11, then a second random sample of size $n_{2}=12$ units will be drawn from that lot and will again be subjected to the test for 1000 hours. Finally, the lot will be accepted if total number of units failed from the two samples are not more than 11, otherwise, the lot will be rejected. Therefore to make a decision whether to accept or reject the proposed lot, an average 43.43 number of units are required under the plan. Next instead of double acceptance sampling plan, if single acceptance sampling plan is to be implemented. Then observe that design parameters from Table \ref{table:3} are as $n=51$ and $c=11$. It emphasizes that a random sample of size $n=51$ units should be subjected to a life test for about 1000 hours. If during the experiment more than 11 units fail then that lot will be rejected, otherwise, it will be accepted. It is seen that $ASN$ obtained using double acceptance sampling plan is smaller than that of the sample size obtained using single acceptance sampling plan. So adopting double acceptance sampling plan is reasonably more economical. Further observe that if at the 5\% producer's risk, the true median life is allowed to increase and set to 6000 , 8000, 10000 and  12000 hours then sample size and acceptance number decrease rapidly as shown in the table below.
\begin{center}
\scriptsize
\begin{tabular}{cccccccccc}
\hline
     &                   & \multicolumn{5}{c}{$\frac{m}{m_{0}}=r_{2}$}\\
     \cline{3-7}
Proposed plan & Design parameter  & 2 & 3 &4 & 5& 6\\
\hline
Double acceptance sampling plan& $(n_{1},n_{2},c_{1},c_{2})$ & (39,12,7,11) & (12,8,0,3) & (9,7,0,2) & (7,6,0,1) & (7,6,0,1)\\
Single acceptance sampling plan &  $(n,c)$                   & (51,11) & (20,3) & (16,2) & (11,1) &(11,1)\\

\hline.
\end{tabular}
\end{center}
\section{\bf{Design of group acceptance sampling plan }}
\label{sec:4}
In this section, group acceptance sampling plan is described:\\

\noindent(1) Draw a random sample of size $n$ units and prefix the number of groups, say $g$. Allocate $r$ \\$~~~~~$ test units to each group so that $n=g \times r$.\\
(2) Select an acceptance number $c$ for a group and put all the groups on the life test for $t_{0}$ units\\$~~~~~$ of time.\\
(3) Accept the lot if at most $c$ number of failures are observed in each group.\\
(4) If in any group $c+1$ or more number of failures are observed then terminate the experiment\\$~~~~~$ and reject the lot.\\

For a given $r$, the proposed group acceptance sampling plan is characterized by two design parameters $(g, c)$. Further for the case $g=1$ the group acceptance sampling plan coincides with the single acceptance sampling plan. Now we observe that the probability of accepting a lot under a group acceptance sampling plan is
\begin{eqnarray}
\label{4.1}
P_{a}(p) &=& \Bigg[\sum_{i=0}^{c}{r \choose i}p^{i}(1-p)^{r-i}\Bigg]^g,
\end{eqnarray}
where $p$ is the probability that a unit fails before the time $t_{0}$ and can be obtained by the equation (\ref{3.2}). Under certain consumer's and producer's risks, together with a quality level, the design parameters $(g,c)$ can be obtained by solving the following optimization problem
\begin{eqnarray*}
\mbox{Minimize} ~~ASN = n= g \times r ~~~~~~~~~~~~~~~~~~~~~~~~~~~~~~~~~~~~~~~~\\
\mbox{Subject to~~~~~~~~~~~~~~~~~~~~~~~~~}~~~~~~~~~~~~~~~~~~~~~~~~~~~~~~~~~~~~~~~\\
P_{a}\Big(p_{1} \mid \frac{m}{m_{0}} = r_{1}\Big) = \Bigg[\sum_{i=0}^{c}{r \choose i}p_{1}^{i}(1-p_{1})^{r-i}\Bigg]^g &\leq& \beta , \\
P_{a}\Big(p_{2} \mid \frac{m}{m_{0}} = r_{2}\Big) = \Bigg[\sum_{i=0}^{c}{r \choose i}p_{2}^{i}(1-p_{2})^{r-i}\Bigg]^g &\geq& 1- \alpha.
\end{eqnarray*}
The probabilities $p_{1}$ and $p_{2}$ are defined in the previous section. Here, it is observed that $ASN$ is given by $n$ and it does not depend on the lot quality. As a consequence, minimizing $ASN$ is equivalent to find the minimum value of $g$ that satisfies the above inequalities for a given $r$.

Next we present the design parameters under group acceptance sampling plan in Table \ref{table:4} for $\gamma=0.75$. These plans are presented for two different choices of $r$ such as $r=5, 10$. From tabulated values, it is observed that a decrease in consumer's risk $\beta$ results an increase in number of groups. Further, as the quality level at producer's risk $r_{2}$ increases, the number of group decreases rapidly. However after a certain level, the probability of accepting a lot starts increasing even though the number of groups and acceptance number are kept at the same level. Effect of the time termination multiplier $a$ can also be seen from the table. For an instance consider the case $r=10$ at quality level $r_{2}=2$, here as $a$ increases, the number of group also increases, whereas, for $r_{2} \geq 3$, the number of groups almost remain the same but acceptance number tend to increase. It is also seen that if an experimenter desires to minimize the total number of units then in that case a different group size may be suggested. As an example, with $\beta=0.25$, $a=0.5$, $r_{2}=2$, we observed that for $r=5$, a total of 471 groups or equivalently a total of 2355 number of units are required to be placed on the life test. However, under the same conditions when $r$ is fixed as 10, then only 24 groups are required or equivalently a total of 240 test units are required to be placed on the test. Therefore, in this case a group of size 10 would be preferred but a group size of 5 would be preferred for $r_{2}=4$. Further Table \ref{table:5}, report design parameters for the shape parameter  value 1.25, that is $\gamma=1.25$. Our reported values reveal that for the associated plan an increase in the value of shape parameter results in a smaller group size. \\

\noindent \textbf{Example 2:}
Assume that lifetime of ball bearings placed on a test follow an inverse Weibull distribution with the shape parameter 0.75 and that the specified life of the ball bearings is 2000 cycles. Further suppose that an experimenter wants to run an experiment for 1000 cycles by implementing 10 units in each group and is interested to know whether median life of the ball bearings is larger than the specified life.  It is further known that the consumer's risk is 10\%, when the true median life is 2000 cycles and the producer's risk is 5\%, when the true median life is 4000 cycles. Thus, we have $\gamma=0.75$, $m=2000$ cycles, $a=0.5$, $r=10$, $\beta=0.10$, $r_{1}=1$, $\alpha=0.05$ and $r_{2}=2$. Then design parameters from Table \ref{table:4} are obtained as $g=40$ and $c=5$. This suggests that a random sample of size 400 units should be drawn and 10 units should be allocated to each of 40 groups. If not more than 5 units fail in each such groups before 1000 cycles, then it will be statistically ensured that the median life of the ball bearings is larger than the specified life. As a consequence, the lot under investigation should be accepted. 
\section{Implementation of sampling plans}
\label{sec:5}
In previous sections we have constructed various sampling plans for a given value of the shape parameter. However, in practice the value of shape parameter for a proposed lot may be unknown. In such situations it is desirable to obtain estimate for the shape parameter before construction of any sampling plan. Sometimes such estimates are obtained using the historical data set or even using a pre-sample for which the corresponding inferences may be available a priori. These statistical inferences can be used to implement a sampling plan for the proposed lot. For simplification we consider a real data set to illustrate  the selection of the shape parameter and the corresponding implementation of sampling plans in practical situations. This data set is given in Lawless \cite{Lawless-2003} and represents the break down times of electrical insulating fluid subject to a 30KV voltage stress. The corresponding break down times (in minutes) are
\begin{verbatim}
7.74, 17.05, 20.46, 21.02, 22.66, 43.40, 47.30, 139.07, 144.12, 175.88, 194.90.
\end{verbatim}
We first verify whether an inverse  Weibull distribution fits the given data set. For comparison purpose, we also consider fitting of  Weibull distribution, lognormal distribution and log-logistic distribution with shape parameter $\gamma$ and scale parameter $\lambda$. We propose to use the method of maximum likelihood for estimating the unknown parameters of each distribution. Further, negative log-likelihood criterion (NLC) and Kolomogorov-Smirnov (K-S) test static values are used to judge the goodness of fit. All the estimated values are reported in Table \ref{table:6}. From tabulated values it is seen that the inverse Weibull distribution fits the given data set reasonably well among other distributions. Therefore the given data set can be analyzed using inverse Weibull distribution.

Now in future suppose that a producer proposes a lot of electrical insulating fluid and claims that the specified median life of the insulation specimens subjected to 30KV voltage stress is 50 minutes. Further consider the consumer's risk 25\% and producer's risk 5\% when the true median life is a multiple of specified median life by 2.0, 3.0 and 4.0. Suppose that an experimenter would like to run the experiment for 25 minutes and is interested to implement double acceptance sampling plan for the proposed lot. In this situation inference available from the given data set is quite useful and accordingly we may select the value of the shape parameter as $\gamma=1.05$, close to the estimated value of shape parameter of the historical data. Consequently by considering $\gamma = 1.05$, we obtain design parameters for the double accepting sampling plans. The values are presented in Table \ref{table:7}. These plans can be used in practice to make adequate inference about the proposed lot. 

Notice that $\gamma$ is unknown for the proposed lot and we considered the estimated value from the historical data set. Therefore it would be further interesting to observe the effect of mis-specification on the lot acceptance probabilities for producer's risk $p_{\alpha}$ and consumer's risk $p_{\beta}$. In consequence assume that the true value of the shape parameter of the proposed lot is $\gamma_{0}$. Now observe that $p$ can be re-expressed as $p_0 = e^{-{(\frac{m}{m_{0}})}^{\gamma_{0}}\,log 2\, a^{-\gamma{0}}}$. Now if design parameters of sampling plans, obtained using $\gamma$, still satisfy the producer's risk and consumer's risk under $p_0$ then the choice of the selected shape parameter $\gamma$ would be considered as reasonably good. For an instance we consider double acceptance sampling plans as reported in Table \ref{table:7} and obtain the acceptance probabilities for different arbitrary choices of $\gamma_{0}$ such as 0.90, 0.95, 1.0, 1.10, 1.15, 1.20. The acceptance probabilities  are reported in Table \ref{table:8}. From the tabulated values it is seen that a higher value of the true shape parameter may satisfies producer's risk and a smaller value may satisfies consumer's risk. This can be treated as a useful information for implementing various sampling plans in practice.
\section{Sampling plans for other percentile points}
\label{sec:6}
So far we have obtained design parameters under double and group acceptance sampling plans when the quality of the test units can be measured by their median lifetime. In this section, we discuss about how these design parameters can also be used further for other percentile points $\theta_{p}$. Assume that for a given plan, a life test is conducted for $\tilde{t_{0}}$ units of time which may be different from the $t_{0}$. Here  $\tilde{t_{0}}=\tilde{a}\theta_{0}$ can be considered as a multiplier of a proposed specified lifetime $\theta_{0}$. Further, let the probability of accepting the lot under the given proposed sampling plan is given by $L_{a}(\tilde{p})$ where $\tilde{p}=F(\tilde{t_{0}}; \gamma, \lambda)=e^{-\frac{\lambda}{\tilde{t_{0}}^{\gamma}}}$. Therefore, using the value of $\lambda$ from expression (\ref{2.2}), $\tilde{p}$ can be written as
\begin{eqnarray}
\label{5.1}
\tilde{p}&=&e^{-\big(\frac{\theta_{p}}{\theta_{0}}\big)^{\gamma}(-log p){\tilde{a}}^{-\gamma}}.
\end{eqnarray}
Consequently, design parameters having minimum sample size, satisfying both the consumer's risk and producer's risk, can be obtained by solving the following optimization problem\\
Minimize $~~ASN$\\
Subject to\\
$~~~~~~~~~~~~~~~~L_{a}(\tilde{p}_{1} \mid \frac{\theta_{p}}{\theta_{0}}=1) \leq \beta $,\\
$~~~~~~~~~~~~~~~~L_{a}(\tilde{p}_{2} \mid \frac{\theta_{p}}{\theta_{0}}=r_{2}^{'}) \geq 1-\alpha$,\\

Also from the equations (\ref{3.2}) and (\ref{5.1}), for given $\alpha, \beta$ and $r_{2}^{'}$ it is observed that sampling plans for $p$th percentile can be obtained using the same table (based on median) for other percentiles provided
\begin{eqnarray}
\label{5.2}
r_{2}=r_{2}^{'} \mbox{~and~}\frac{m}{m_{0}}=\frac{\theta_{p}}{\theta_{0}}~~\mbox{or}~~ r_{2}=r_{2}^{'} ~ \mbox{and} ~
a=\Bigg(\tilde{a}\Big(-\frac{log 2}{log p}\Big)^{\frac{1}{\gamma}}\Bigg).
\end{eqnarray}
For instance consider Example 1 where an experimenter wishes to obtain sampling plans for 75th percentile point instead of median lifetime. It is known that the corresponding experiment has been performed for 620 hours. Further assume that values of $\beta, \alpha$ and $r_2$ remain the same. So, we have $\tilde{a}=0.31$ and using equation (\ref{5.2}) we find that $a=1.0$. Therefore sampling plan based on medians corresponding to $a=1.0$ can be used in practice.
\section{Conclusion}
\label{sec:7}
In this paper, double acceptance sampling plan and group acceptance sampling plan are proposed for inverse Weibull distribution based on truncated life tests under the assumption that the median lifetime is a quality parameter. Design parameters for various sampling plans are obtained under certain consumer's risk and producer's risk involved simultaneously. Proposed plans are explained with the help of examples and tables. We have also observed that tables based on medians can be used for other percentiles points as well. Finally, we have analyzed a real data set to illustrate the selection of shape parameter and implementation of sampling plans in practical situations. The case of mis-specification of shape parameter has also been discussed. It is seen that double acceptance sampling plans are more economical than single acceptance sampling plan except possibly zero acceptance sampling plan. We have also observed that group acceptance sampling plans with different number of units in each group may reduce the sample size. Finally we mention that a search for algorithm is required in order to construct a sampling plan which provide time termination multiplier, group size with smaller sample size under some cost constraints. This may be a future direction. 

\newpage
\normalsize
\begin{table}[t]
\tabcolsep 4.2pt
\begin{center}
\caption{Double acceptance sampling plan for $\gamma = 0.75$}
\label{table:1}
\medskip
\scriptsize
\begin{tabular}{cccccccccccccccccccccc}
\toprule
           &         &  \multicolumn{6}{c}{$a=0.5$} &&  \multicolumn{6}{c}{$a=0.7$} &&  \multicolumn{6}{c}{$a=1.0$} \\ 
           \cline{3-8} \cline{10-15} \cline{17-22}
 $\beta$  & $r_{2}$ & $n_{1}$&$n_{2}$ & $c_{1}$&$c_{2}$ & $ASN$ & $p_{\alpha}$ && $n_{1}$&$n_{2}$ & $c_{1}$&$c_{2}$ & $ASN$ & $p_{\alpha}$ && $n_{1}$&$n_{2}$ & $c_{1}$&$c_{2}$ & $ASN$ & $p_{\alpha}$ \\
\midrule
  0.25 & 2 & 23& 7 & 4 & 7 & 26.21 & 0.9510 && 27&  7& 8& 11 &29.96 & 0.9557 && 30& 8 & 12& 16 & 34.21 & 0.9534\\
       & 3 &  7& 5 & 0 & 2 & 9.73  & 0.9548 && 10&  5& 2&  4 &12.30 & 0.9716 && 11& 3 &  3&  5 & 12.16 & 0.9521\\
       & 4 &  5& 4 & 0 & 1 & 6.39  & 0.9650 &&  5&  4& 0&  2 &7.40  & 0.9741 &&  7& 4 &  2&  3 &  8.09 & 0.9661\\
       & 5 &  5& 4 & 0 & 1 & 6.39  & 0.9887 &&  4&  2& 0&  1 &4.68  & 0.9711 &&  4& 3 &  0&  2 &  5.87 & 0.9759\\
       & 6 &  5& 4 & 0 & 1 & 6.39  & 0.9962 &&  4&  2& 0&  1 &4.68  & 0.9875 &&  3& 2 &  0&  1 &  3.75 & 0.9613\\
  0.10 & 2 & 39& 12& 7 &11 & 43.43 & 0.9552 && 41& 11& 9& 16 &46.33 & 0.9540 && 50& 9 & 18& 24 & 53.70 & 0.9550\\
       & 3 & 12& 8 & 0 & 3 & 15.56 & 0.9531 && 14&  7& 2&  5 &17.04 &0.9604  && 15& 6 &  3&  7 & 17.89 & 0.9502\\
       & 4 &  9& 7 & 0 & 2 & 11.78 & 0.9814 && 10&  5& 1&  3 &11.64 & 0.9773 &&  9& 5 &  1&  4 & 11.40 & 0.9639\\
       & 5 &  7& 6 & 0 & 1 & 8.39  & 0.9774 &&  7&  5& 0&  2 &8.92  & 0.9828 &&  5& 4 &  0&  2 &  6.87 & 0.9511\\
       & 6 &  7& 6 & 0 & 1 & 8.39  & 0.9923 &&  5&  5& 0&  1 &6.27  & 0.9710 &&  5& 4 &  0&  2 &  6.87 & 0.9799\\
  0.05 & 2 & 54& 9 & 10&13 & 55.22 & 0.9502 && 53& 14&12& 20 &58.54 & 0.9554 && 64& 12& 24& 30 & 67.88 & 0.9557\\
       & 3 & 17& 10& 0 & 4 & 20.47 & 0.9626 && 18&  9& 2&  6 &21.17 & 0.9537 && 20& 8 &  5&  9 & 23.13 & 0.9557\\
       & 4 & 11& 8 & 0 & 2 & 13.13 & 0.9701 && 10&  7& 0&  3 &12.56 & 0.9630 && 12& 6 &  1&  5 & 14.30 & 0.9683\\
       & 5 &  9& 6 & 0 & 1 & 9.84  & 0.9686 &&  8&  6& 0&  2 &9.74  & 0.9737 &&  8& 5 &  0&  3 &  9.79 & 0.9676\\
       & 6 &  9& 6 & 0 & 1 & 9.84  & 0.9892 &&  7&  4& 0&  1 &7.50  & 0.9604 &&  6& 5 &  0&  2 &  7.64 & 0.9648\\
 0.01  & 2 & 70& 26& 11&19 & 77.17 & 0.9563 && 76& 22&15& 28 &82.68 & 0.9560 && 88& 19& 29& 41 & 93.62 & 0.9536\\
       & 3 & 22& 16& 0 & 5 & 26.35 & 0.9527 && 25& 13& 2&  8 &28.36 & 0.9554 && 30& 12&  6& 13 & 33.49 & 0.9627\\
       & 4 & 15& 15& 0 & 3 & 18.90 & 0.9772 && 14& 11& 0&  4 &16.94 & 0.9607 && 15& 10&  2&  6 & 17.99 & 0.9502\\
       & 5 & 14& 12& 0 & 2 & 15.60 & 0.9859 && 11& 11& 0&  3 &14.11 & 0.9809 && 11& 8 &  0&  4 & 13.19 & 0.9669\\
       & 6 & 13& 9 & 0 & 1 & 13.41 & 0.9778 && 10&  9& 0&  2 &11.39 & 0.9816 &&  9& 8 &  0&  3 & 11.01 & 0.9732\\
\bottomrule
\end{tabular}
\end{center}
\end{table}

\normalsize
\begin{table}[p]
\tabcolsep 4.2pt
\begin{center}
\caption{Double acceptance sampling plan for $\gamma = 1.25$}
\label{table:2}
\medskip
\scriptsize
\begin{tabular}{cccccccccccccccccccccc}
\toprule
           &         &  \multicolumn{6}{c}{$a=0.5$} &&  \multicolumn{6}{c}{$a=0.7$} &&  \multicolumn{6}{c}{$a=1.0$} \\ 
           \cline{3-8} \cline{10-15} \cline{17-22}
 $\beta$  & $r_{2}$ & $n_{1}$&$n_{2}$ & $c_{1}$&$c_{2}$ & $ASN$ & $p_{\alpha}$ && $n_{1}$&$n_{2}$ & $c_{1}$&$c_{2}$ & $ASN$ & $p_{\alpha}$ && $n_{1}$&$n_{2}$ & $c_{1}$&$c_{2}$ & $ASN$ & $p_{\alpha}$ \\
 \midrule
0.25  & 2 & 8  &8  & 0& 1 & 10.76 & 0.9694 &&  6& 5 & 0& 2  & 8.93 & 0.9566 && 11& 3 & 3& 5 & 12.16 & 0.9647\\
      & 3 & 8  &8  & 0& 1 & 10.76 & 0.9998 &&  5& 3 & 0& 1  & 5.97 & 0.9954 &&  3& 2 & 0& 1 &  3.75 & 0.9666\\
      & 4 & 8  &8  & 0& 1 & 10.76 &1       &&  5& 3 & 0& 1  & 5.97 & 0.9999 &&  3& 2 & 0& 1 &  3.75 & 0.9966\\
      & 5 & 8  &8  & 0& 1 & 10.76 &1       &&  5& 3 & 0& 1  & 5.97 & 1      &&  3& 2 & 0& 1 &  3.75 & 0.9998\\
      & 6 & 8  &8  & 0& 1 & 10.76 &1       &&  5& 3 & 0& 1  & 5.97 & 1      &&  3& 2 & 0& 1 &  3.75 & 1\\
 0.10 & 2 & 16 &11 & 0& 2 & 19.82 & 0.9849 && 11& 7 & 0& 3  & 14.12& 0.9570 && 15& 6 & 3& 7 & 17.89 & 0.9655\\
      & 3 & 12 &11 & 0& 1 & 14.42 & 0.9996 &&  7& 4 & 0& 1  &7.79  & 0.9912 &&  5& 4 & 0& 2 &  6.87 & 0.9837\\
      & 4 & 12 &11 & 0& 1 & 14.42 &1       &&  7& 4 & 0& 1  &7.79  & 0.9998 &&  4& 3 & 0& 1 &  4.75 & 0.9933\\
      & 5 & 12 &11 & 0& 1 & 14.42 &1       &&  7& 4 & 0& 1  &7.79  & 1      &&  4& 3 & 0& 1 &  4.75 & 0.9994\\
      & 6 & 12 &11 & 0& 1 & 14.42 &1       &&  7& 4 & 0& 1  &7.79  &  1     &&  4& 3 & 0& 1 &  4.75 & 1     \\
 0.05 & 2 & 18 &15 & 0& 2 & 22.15 & 0.9748 && 15&10 & 1& 4  & 18.69& 0.9643 && 18& 8 & 4& 8 & 21.13 & 0.9544\\
      & 3 & 15 &13 & 0& 1 & 16.88 & 0.9993 &&  8& 6 & 0& 1  & 8.89 & 0.9867 &&  6& 5 & 0& 2 &  7.64 & 0.9714\\
      & 4 & 15 &13 & 0& 1 & 16.88 & 1      &&  8& 6 & 0& 1  & 8.89 & 0.9996 &&  5& 4 & 0& 1 &  5.62 & 0.9892\\
      & 5 & 15 &13 & 0& 1 & 16.88 &1       &&  8& 6 & 0& 1  & 8.89 & 1      &&  5& 4 & 0& 1 &  5.62 & 0.9991\\
      & 6 & 15 &13 & 0& 1 & 16.88 &1       &&  8& 6 & 0& 1  & 8.89 & 1      &&  5& 4 & 0& 1 &  5.62 & 1     \\
 0.01 & 2 & 26 &16 & 0& 2 & 27.52 & 0.9517 && 20&15 & 0& 5  &24.19 & 0.9532 && 26& 12& 6&11 & 29.28 & 0.9551\\
      & 3 & 22 &19 & 0& 1 & 22.90 & 0.9986 && 12& 7 & 0& 1  &12.30 & 0.9749 &&  9& 8 & 0& 3 & 11.01 & 0.9793\\
      & 4 & 22 &19 & 0& 1 & 22.90 & 0.9999 && 12& 7 & 0& 1  &12.30 & 0.9993 &&  7& 5 & 0& 1 &  7.27 & 0.9805\\
      & 5 & 22 &19 & 0& 1 & 22.90 & 1      && 12& 7 & 0& 1  &12.30 & 0.9999 &&  7& 5 & 0& 1 &  7.27 & 0.9983\\
      & 6 & 22 &19 & 0& 1 & 22.90 & 1      && 12& 7 & 0& 1  &12.30 & 1      &&  7& 5 & 0& 1 &  7.27 & 0.9999\\

\bottomrule
\end{tabular}
\end{center}
\end{table}
\normalsize
\begin{table}[p]
\tabcolsep 4.2pt
\begin{center}
\caption{Comparison between single and double acceptance sampling plans}
\label{table:3}
\medskip
\scriptsize
\begin{tabular}{ccccccccccccccccccccccc}
\toprule
       &       && \multicolumn{8}{c}{$\gamma=0.75$} &&& \multicolumn{8}{c}{$\gamma=1.25$}\\
       \cline{4-11} \cline{14-21}
       &      && \multicolumn{2}{c}{$a=0.5$} && \multicolumn{2}{c}{$a=0.7$} && \multicolumn{2}{c}{$a=1.0$} &&& \multicolumn{2}{c}{$a=0.5$}&&\multicolumn{2}{c}{$a=0.7$}&&\multicolumn{2}{c}{$a=1.0$}\\
       \cline{4-5}\cline{7-8}\cline{10-11}\cline{14-15}\cline{17-18}\cline{20-21}
$\beta$&$r_{2}$  && ADP & SSP && ADP & SSP &&ADP &SSP &&& ADP &SSP &&ADP &SSP &&ADP &ASP\\
\midrule
0.25   & 2 && 26.21 & 34(8) &&29.96 & 36(12) && 34.21 & 40(17) &&& 10.76 &14(1)&& 8.93 &11(2)&& 12.16 &14(5)\\
       & 3 && 9.73  & 12(2) &&12.30 & 15(4)  &&12.16  & 14(5)  &&&10.76 &7(0) && 5.97 &8(1) && 3.75 &5(1)\\
       & 4 && 6.39  & 8(1)  &&7.40  & 9(2)   && 8.09  & 10(3)  &&&10.76 &7(0) &&5.97 &4(0) && 3.75 &2(0)\\
       & 5 && 6.39  & 8(1)  &&4.68  &6(1)&& 5.87  & 7(2)   &&&10.76 &7(0) &&5.97 &4(0) && 3.75 &2(0)\\
       & 6 &&6.39  & 4(0)  &&4.68  &6(1)&& 3.75  & 5(1)   &&&10.76 &7(0) &&5.97 &4(0) && 3.75 &2(0)\\
  0.10 & 2 && 43.43 & 51(11)&&46.33 & 52(16) && 53.70 & 59(24) &&& 19.82 &26(2)&& 14.12 &18(3)&& 17.89 &21(7)\\
       & 3 && 15.56 & 20(3) &&17.04 & 21(5)  && 17.89 & 24(8)  &&& 14.42 &11(0)&&7.79  &10(1)&& 6.87 &9(2)\\
      & 4 &&  11.78 & 16(2) &&11.64 & 15(3)  && 11.40 & 14(4)  &&& 14.42 &11(0)&&7.79  &6(0) && 6.87 &7(1)\\
      & 5 &&  8.39  & 11(1) &&8.92  & 12(2)  && 6.87  & 12(3)  &&& 14.42 &11(0)&&7.79  &6(0) && 6.87 &4(0)\\
      & 6 &&  8.39  & 11(1) && 6.27 & 9(1)   &&6.87  & 9(2)   &&& 14.42 &11(0)&&7.79  &6(0) && 6.87 &4(0)\\
  0.05 & 2 && 55.22 & 66(14)&& 58.54 & 67(20) && 67.88 & 76(30)&&& 22.15 &31(2)&& 18.69 &25(4)&& 21.13 &26(8)\\
       & 3 && 20.47 & 27(4) && 21.17 & 27(6)  &&23.13  & 28(9) &&& 16.88 &15(0)&& 8.89 &12(1)&&7.64   &11(2)\\
       & 4 && 13.13 & 18(2) &&12.56  & 17(3)  && 14.30 & 18(5) &&& 16.88 &15(0)&& 8.89 &8(0) &&7.64   &8(1)\\
      & 5  && 9.84 &  14(1) && 9.74  & 14(2)  &&10.29 & 13(3)  &&& 16.88 &15(0)&& 8.89 &8(0) &&7.64   &5(0)\\
      & 6 && 9.84 &  14(1) && 7.50  & 10(1)  && 7.64  & 11(2) &&& 16.88 &15(0)&& 8.89 &8(0) &&7.64   &5(0)\\
 0.01 & 2 &&  77.17 & 96(19)&&82.68& 98(28) &&93.62& 107(41)&&&27.52 &41(2)&&24.19 &35(5)&&29.28 &38(11)\\
      & 3 &&  26.35 & 38(5) &&28.36& 38(8)  &&33.49& 42(13) &&&22.90 &22(0)&&12.30 &17(1)&&11.01 &17(3)\\
      & 4 &&  18.90 & 29(3) &&16.94& 25(4)  &&17.99& 27(7)  &&&22.90 &22(0)&&12.30 &12(0)&&7.27 &11(1)\\
      & 5 &&  15.60 & 24(2) &&14.11& 21(3)  &&13.19& 19(4)  &&&22.90 &22(0)&&12.30 &12(0)&&7.27 &7(0)\\
      & 6&&   13.41 & 19(1) &&11.39& 18(2)  &&11.01& 17(3)  &&&22.90 &22(0)&&12.30 &12(0)&&7.27 &7(0)\\
\bottomrule
\end{tabular}\\
Note: ADP - Average sample number of double acceptance sampling plan.\\
SSP - Sample size of single acceptance sampling plan (acceptance number).\\
\end{center}
\end{table}
\normalsize
\begin{table}[p]
\tabcolsep 4.2pt
\begin{center}
\caption{Group acceptance sampling plan for $\gamma = 0.75$}
\label{table:4}
\medskip
\scriptsize
\begin{tabular}{cccccccccccccccccccccc}
\toprule
        &       &  \multicolumn{9}{c}{$r=5$} && \multicolumn{9}{c}{$r=10$}\\ \cline{3-11} \cline{13-21}
$\beta$&$r_{2}$& \multicolumn{3}{c}{$a=0.5$}& \multicolumn{3}{c}{$a=0.7$} & \multicolumn{3}{c}{$a=1.0$} && \multicolumn{3}{c}{$a=0.5$}& \multicolumn{3}{c}{$a=0.7$} & \multicolumn{3}{c}{$a=1.0$}\\
       &        &$g$ & $c$ &$p_{\alpha}$&$g$ & $c$ &$p_{\alpha}$&$g$ & $c$ &$p_{\alpha}$ &&$g$ & $c$ &$p_{\alpha}$&$g$ & $c$ &$p_{\alpha}$&$g$ & $c$ &$p_{\alpha}$ \\
\midrule
0.25 & 2 &471&4 &0.9743 &-  &-&-&-&-&-&&24 &5 &0.9767 &24 &6 &0.9648 &129&8&0.9745\\
     &3 & 8 & 2 &0.9755 & 15 &3 &0.9827 &44 &4&0.9838 && 3 &3 &0.9892&3&4&0.9858&3&5&0.9779\\
      & 4 &  3&1 &0.9624 &4 &2 &0.9838 &2 &2 &0.9558&&2  &2 &0.9900 &1  &2 &0.9635 &1 &3 &0.9593\\
      & 5 &  3&1 &0.9882& 2 &1& 0.9580&2 &2&0.9837&&1 & 1&0.9834&1&2&0.9895&1 &3&0.9879\\
      & 6 &  3&1 &0.9961 &2 &1 &0.9819 &1 &1 &0.9573&&1  &1 &0.9944 &1  &1 &0.9632 &1 &2 &0.9715\\
 0.10 & 2 &782&4 &0.9577 & - &-  &-   & - & - &-&&40 &5 &0.9615 &173&7&0.9745&214&8&0.9581\\
      & 3& 12 & 2 &0.9634 & 25 & 3 &0.9713& 73 &4&0.9733&&5 & 3 &0.9821&5&4&0.9764&5&5&0.9635\\
      & 4 &  4&1 &0.9502 &6 &2 &0.9758 &12&3 &0.9793&&3  &2 &0.9851 &3  &3 &0.9848 &3 &4 &0.9776\\
      & 5&  4&1 &0.9843&6&2&0.9937&4 &2&0.9676&&2 &1 &0.9670&2 &2 &0.9791&2 &3&0.9759\\
      & 6 &  4&1 &0.9949 &3 &1 &0.9731 &4 &2 &0.9877&&2  &1 &0.9889 &1  &1 &0.9632 &1 &2 &0.9715\\
 0.05 & 2 & -  & - & -  & - & - & -     & - & - &-&&52 &5 &0.9503 &226&7&0.9668&3067&9&0.9738\\
      & 3 & 16 & 2 & 0.9516 &32 &3 &0.9634& 95 &4&0.9654&&7 &3&0.9750&7&4&0.9672&16&6&0.9834\\
      & 4 &16 &2&0.9924 &8 &2 &0.9678 &15&3 &0.9742&&3  &2 &0.9851 &4  &3 &0.9798 &4 &4 &0.9703\\
      & 5 & 5 & 1 & 0.9804&8&2&0.9916&5&2&0.9597&&2 &1&0.9670&2 &2&0.9791&2&3&0.9759\\
      & 6 &5  &1&0.9936 &3 &1 &0.9731 &5 &2 &0.9846&&2  &1 &0.9889 &1  &1 &0.9632 &2 &3 &0.9928\\
 0.01 & 2 &  -   & - & -  &-  & - & - &-&-&-&& 345&6&0.9698&2509&8&0.9778&4714&9&0.9600\\
      & 3 & 128 &3&0.9855& 425 &4&0.9861 &-& - &-&&10&3&0.9645&10&4&0.9534&25&6&0.9743\\
      & 4 & 24&2 &0.9886 &12 &2 &0.9521 &23 &3 &0.9607&&5  &2 &0.9753 &5  &3 &0.9748 &5  &4 &0.9630\\
      & 5&7 & 1 & 0.9727 &12 & 2&0.9874 &23 & 3& 0.9900&&3&1&0.9510&3 &2&0.9688&3 &3&0.9640\\
      & 6 &7  &1 &0.9910 &5  &1 &0.9555 &7  &2 &0.9785&&3  &1 &0.9834 &3  &2 &0.9909 &3  &3 &0.9892\\
\bottomrule
\end{tabular}\\
Note: (-) indicates that no $g$ and $c$ could satisfy the conditions.\\
\end{center}
\end{table}
\normalsize
\begin{table}[p]
\tabcolsep 4.2pt
\begin{center}
\caption{Group acceptance sampling plan for $\gamma = 1.25$}
\label{table:5}
\medskip
\scriptsize
\begin{tabular}{cccccccccccccccccccccc}
\toprule
        &       &  \multicolumn{9}{c}{$r=5$} && \multicolumn{9}{c}{$r=10$}\\ \cline{3-11} \cline{13-21}
$\beta$&$r_{2}$& \multicolumn{3}{c}{$a=0.5$}& \multicolumn{3}{c}{$a=0.7$} & \multicolumn{3}{c}{$a=1.0$} && \multicolumn{3}{c}{$a=0.5$}& \multicolumn{3}{c}{$a=0.7$} & \multicolumn{3}{c}{$a=1.0$}\\
       &        &$g$ & $c$ &$p_{\alpha}$&$g$ & $c$ &$p_{\alpha}$&$g$ & $c$ &$p_{\alpha}$ &&$g$ & $c$ &$p_{\alpha}$&$g$ & $c$ &$p_{\alpha}$&$g$ & $c$ &$p_{\alpha}$ \\
\midrule
0.25&2&5&1&0.9813&6&2&0.9766&7&3&0.9602&&2&1&0.9684&3&3&0.9855&3&5&0.9845\\
&3&2&0&0.9852&2&1&0.9962&1&1&0.9632&&1&0&0.9852&1&1&0.9919&1&2&0.9769\\
&4&2&0&0.9991&1&0&0.9891&1&1&0.9962&&1&0&0.9991&1&0&0.9783&1&1&0.9841\\
&5&2&0&0.9999&1&0&0.9985&1&0&0.9722&&1&0&0.9999&1&0&0.9969&1&1&0.9986\\
&6&2&0&1&1&0&0.9998&1&0&0.9926&&1&0&1&1&0&0.9996&1&0&0.9852\\
0.10&2&9&1&0.9665&10&2&0.9614&73&4&0.9810&&3&1&0.9530&4&3&0.9807&5&5&0.9743\\
&3&3&0&0.9779&3&1&0.9943&4&2&0.9902&&2&0&0.9706&1&1&0.9919&1&2&0.9769\\
&4&3&0&0.9987&2&0&0.9783&2&1&0.9925&&2&0&0.9982&1&0&0.9783&1&1&0.9841\\
&5&3&0&0.9999&2&0&0.9969&1&0&0.9722&&2&0&0.9999&1&0&0.9969&1&1&0.9986\\
&6&3&0&1&2&0&0.9996&1&0&0.9926&&2&0&1&1&0&0.9996&1&0&0.9852\\
0.05&2&11&1&0.9592&13&2&0.9510&95&4&0.9753&&9&2&0.9924&5&3&0.9759&7&5&0.9642\\
&3&3&0&0.9779&4&1&0.9925&5&2&0.9877&&2&0&0.9706&2&1&0.9839&2&2&0.9543\\
&4&3&0&0.9987&2&0&0.9783&2&1&0.9925&&2&0&0.9982&1&0&0.9783&1&1&0.9841\\
&5&3&0&0.9999&2&0&0.9969&1&0&0.9722&&2&0&0.9999&1&0&0.9969&1&1&0.9986\\
&6&3&0&1&2&0&0.9996&1&0&0.9926&&2&0&1&1&0&0.9996&1&0&0.9852\\
0.01&2&86&2&0.9935&94&3&0.9852&146&4&0.9623&&13&2&0.9891&8&3&0.9618&25&6&0.9833\\
&3&5&0&0.9634&6&1&0.9887&7&2&0.9829&&3&0&0.9562&2&1&0.9839&2&2&0.9543\\
&4&5&0&0.9978&3&0&0.9676&3&1&0.9887&&3&0&0.9973&2&0&0.9571&2&1&0.9684\\
&5&5&0&0.9999&3&0&0.9954&3&1&0.9991&&3&0&0.9999&2&0&0.9939&2&1&0.9972\\
&6&5&0&1&3&0&0.9994&2&0&0.9852&&3&0&1&2&0&0.9992&1&0&0.9852\\
\bottomrule
\end{tabular}
\end{center}
\end{table}

\normalsize
\begin{table}[p]
\tabcolsep 4.2pt
\begin{center}
\caption{Goodness of fit tests for given distributions}
\label{table:6}
\medskip
\scriptsize
\begin{tabular}{ccccc}
\toprule
Model & $\hat{\gamma}$ & $\hat{\lambda}$ & NLC & K-S\\
\midrule
Inweibull   & 1.05411&32.3524&58.535&0.2004\\
Weibull     & 1.05881&0.01288&58.578&0.2166\\
Lognormal   & 3.82199&1.05948&58.285&0.2168\\
Log-logistic& 3.79847&0.65422&58.853&0.2146\\
\bottomrule
\end{tabular}
\end{center}
\end{table}

\normalsize
\begin{table}[p]
\tabcolsep 4.2 pt
\begin{center}
\caption{{Double acceptance sampling plans for the given data set}}
\label{table:7}
\medskip
\scriptsize
  \begin{tabular}{ccccccccccc}
  \toprule
$\beta$&$r_{2}$&$n_{1}$&$n_2$&$c_1$&$c_2$&$ASN$&$p_{\alpha}$\\
\toprule
0.25&2.0& 9 & 7& 0& 2& 12.83& 0.9568\\
	&3.0& 7 & 5& 0& 1& 8.63 & 0.9941\\
	&4.0& 7 & 5& 0& 1& 8.63 & 0.9997\\
\bottomrule
\end{tabular}
\end{center}
\end{table}

\begin{table}[p]
\tabcolsep 4.2 pt
\begin{center}
\caption{{Acceptance probabilities under misspecification of shape parameter $\gamma$}}
\label{table:8}
\medskip
\scriptsize
  \begin{tabular}{ccccccccccccccccccccccccc}
  \toprule
   &       &\multicolumn{2}{c}{$\gamma_0=0.90$}&\multicolumn{2}{c}{$\gamma_0=0.95$}&\multicolumn{2}{c}{$\gamma_0=1.0$}&\multicolumn{2}{c}{$\gamma=1.05$}&\multicolumn{2}{c}{$\gamma_0=1.10$}&\multicolumn{2}{c}{$\gamma_0=1.15$}&\multicolumn{2}{c}{$\gamma_0=1.20$}\\
   \cline{3-4}\cline{7-8}\cline{11-12}\cline{15-16}
$\beta$&$r_{2}$&$p_{\beta}$&$p_{\alpha}$&$p_{\beta}$&$p_{\alpha}$&$p_{\beta}$&$p_{\alpha}$&$p_{\beta}$&$p_{\alpha}$
  &$p_{\beta}$&$p_{\alpha}$&$p_{\beta}$&$p_{\alpha}$&$p_{\beta}$&$p_{\alpha}$\\
\toprule
0.25&2.0&0.1596&0.8480& 0.1860&0.8915& 0.2154&0.9297& 0.2475&0.9568& 0.2823&0.9750& 0.3196&0.9863& 0.3591&0.9929\\
 &3.0&0.1624&0.9558& 0.1839&0.9757& 0.2073&0.9875& 0.2327&0.9941& 0.2601&0.9974& 0.2892&0.9989& 0.3201&0.9996\\
 &4.0&0.1624&0.9936& 0.1839&0.9975& 0.2073&0.9991& 0.2321&0.9997& 0.2601&0.9999& 0.2892&0.9999& 0.3201&0.9999\\
\bottomrule
\end{tabular}
\end{center}
\end{table}


\begin{thebibliography}{ygsh}
\bibitem{Epstein-1954}
Epstein, B.
Truncated life tests in the exponential case.
{\it Annals of Mathematical Statistics}. 1954; 25(3): 555-564.

\bibitem{Goode-Kao-1961}
Goode, H.P. and Kao, J.H.K.
Sampling plans based on the Weibull distribution.
{\it Proceeding of the Seventh National Symposium on Reliability and Quality Control}.  Philadelphia. 1961; 24-40.

\bibitem{Gupta-1962}
Gupta, S.S.
Life test sampling plans for normal and lognormal distributions.
{\it Technometrics}. 1962; 4(2): 151-175.

\bibitem{Rosaiah-Kantam-2005}
Rosaiah, K. and Kantam, R.R.L.
Acceptance sampling based on the inverse Rayleigh distribution.
{\it Economic Quality Control}. 2005; 20(2): 277-286.

\bibitem{Rosaiah-Kantam-Kumar-2006}
Rosaiah, K., Kantam, R.R.L. and Kumar, Ch.S.
Reliability of test plans for exponentiated log-logistic distribution.
{\it Economic Quality Control}. 2006; 21(2): 165-175.

\bibitem{Tsai-Wu-2006}
Tsai, T.R. and Wu, S.J.
Acceptance sampling based on truncated life tests for generalized Rayleigh distribution.
{\it Journal of Applied Statistics}. 2006; 33(6): 595-600.

\bibitem{Balakrishnan-Leiva-Lopez-2007}
Balakrishnan, N., Leiva, L. and Lopez, J.
Acceptance sampling plans from truncated life tests based on the generalized Birnbaum-Saunders distribution.
{\it Communications in Statistics - Simulation and Computation}. 2007; 36: 643-656.

\bibitem{Aslam-Kundu-Ahmed-2010}
Aslam, M., Kundu, D. and Ahmad, M.
Time truncated acceptance sampling plans for generalized exponential distribution.
{\it Journal of Applied Statistics}. 2010; 37(4): 555-566.

\bibitem{Aslam-Jun-2010}
Aslam, M. and Jun, C.H.
A double acceptance sampling plans for generalized log-logistic distribution with known shape parameter.
{\it Journal of Applied Statistics}, 2010; 37(3) : 405-414.

\bibitem{Aslam-Jun-Ahmed-2011}
Aslam, M., Jun, C.H. and Ahmed, M.
New acceptance sampling plans based on life tests for Birnbaum-Saunders distributions.
{\it Journal of Statistical Computation and Simulation}. 2011; 81(4): 461-470.

\bibitem{Aslam-Jun-Ahmed-2009}
Aslam, M., Jun, C.H. and Ahmad, M.
A Group sampling plan based on truncated life test for Gamma distributed items.
{\it Pakistan Journal of Statistics}. 2009; 25(3): 333-340.

\bibitem{Aslam-Junn-2009}
Aslam, M. and Jun, C.H.
A group acceptance sampling plans for truncated life test having Weibull distribution.
{\it Journal of Applied Statistics}. 2009; 36(9): 1021-1027.

\bibitem{Aslam-Jun-2009}
Aslam, M. and Jun, C.H.
A group acceptance sampling plans for truncated life tests based on the inverse Rayleigh and log-logistic distributions.
{\it Pakistan Journal of Statistics}. 2009; 25(2): 107-119.

\bibitem{singh2015sampling}
Singh, S., Tripathi, Y.M. and Jun, C.H.
Sampling plans based on truncated life test for a generalized inverted exponential distribution.
{\it Industrial Engineering \& Management Systems}. 2015; 14(2):183-195.

\bibitem{Keller-Kanath-1982}
Killer, A.Z. and Kamath, A.R.R.
Alternative reliability models for mechanical systems.
{\it Third International Conference on Reliability and Maintainability 18-21 October}, Toulse, France. 1982.

\bibitem{Kundu-Howlader-2010}
Kundu, D. and Howlader, H.
Bayesian inference and prediction of the inverse Weibull distribution for Type-II censored data.
{\it Computational Statistics and Data Analysis}. 2010; 54: 1547-1558.

\bibitem{Erto-Rapone-1984}
Erto, P. and Rapone, M.
Non-informative and practical Bayesian confidence bounds for reliable life in the Weibull model.
{\it Reliability Engineering}. 1984; 7: 181-191.

\bibitem{Calabria-Pulcini-1994}
Calabria, R. and Pulcini, G.
Bayesian 2-sample prediction for the inverse Weibull distribution.
{\it Communications in Statistics - Theory and Methods}. 194; 23: 1811-1824.

\bibitem{Murthy-2004}
Murthy, D.N.P., Xie, M. and Jiang, R.
Weibull Models. Wiley, New York 2004.

\bibitem{Nelson-1982}
Nelson, W.
{\it Applied Life Data Analysis}, Wiley. 1982.

\bibitem{Erto-1989}
Erto, P.
Genesis properties and identification of the inverse Weibull lifetime model.
{\it Statitica Applicata}. 1989;  1: 117-128.

\bibitem{Kim-Lee-Kang-2012}
Kim, D.H., Lee, W.D. and Kang, S.G.
Non-informative priors for the inverse Weibull distribution.
{\it Journal of Statistical Computation and Simulation}. 2012; {doi. 10.1080/00949655.2012.739171}.

\bibitem{Sultan-Alsadat-Kundu-2013}
Sultan, K.S., Alsadat, N.H. and Kundu, D.
Bayesian and maximum likelihood estimations of the inverse Weibull parameters under progressive type-II censoring.
{\it Journal of Statistical Computation and Simulation}. 2013; {doi. 10.1080/00949655.2013.788652}.

\bibitem{Lawless-2003}
Lawless, J., Statistical Models and Methods for Lifetime Data. New York, Wiley, 2003.

\end{thebibliography}
\end{document}